\documentclass[journal=jacsat,manuscript=article]{achemso}

\usepackage[version=3]{mhchem} 
\usepackage[T1]{fontenc}       

\usepackage{graphicx}
\usepackage[mediumspace,mediumqspace,squaren]{SIunits}
\usepackage{subcaption}

\author{Vishal Vashistha}
\affiliation{Faculty of Physics, Adam Mickiewicz University in Poznan, Poland}
\altaffiliation{these authors contributed equally to this work.}
\email{visvas@amu.edu.pl}
\author{Gayatri Vaidya}
\affiliation{Centre of Excellence in Nanoelectronics - CEN, IIT Bombay, India 400076}
\altaffiliation{these authors contributed equally to this work.}
\author{Ravi S. Hegde}
\affiliation{Indian Institute of Technology, Gandhinagar, India 382355}
\author{Andriy E Serebryannikov}
\affiliation{Faculty of Physics, Adam Mickiewicz University in Poznan, Poland}
\author{Nicolas Bonod}
\affiliation{Aix Marseille Univ, CNRS, Centrale Marseille, Institut Fresnel, 13013 Marseille, France}

\author{Maciej Krawczyk}
\email{krawczyk@amu.edu.pl}
\affiliation{Faculty of Physics, Adam Mickiewicz University in Poznan, Poland}

\setkeys{acs}{articletitle=true}
\setkeys{acs}{keywords=true}

\title[An \textsf{achemso} demo]
{All-Dielectric Metasurfaces Based on Cross-Shaped Resonators for Color Pixels with Extended Gamut}


\begin{document}

\begin{abstract}
Printing technology based on plasmonic structures has many advantages over pigment based color printing such as high resolution, ultra-compact size and low power consumption. However, due to high losses and broad resonance behavior of metals in the visible spectrum, it becomes challenging to produce well-defined colors. Here, we investigate cross-shaped dielectric nanoresonators which enable high quality resonance in the visible spectral regime and, hence, high quality colors. We numerically predict and experimentally demonstrate that the proposed all-dielectric nanostructures exhibit 
high quality colors with selective wavelengths, in particular, due to 
lower losses as compared to metal based plasmonic filters. This results in fundamental colors (RGB) with high hue and saturation. We further show that a large gamut of colors can be achieved by selecting the appropriate length and width of individual $Si$ nanoantennas. Moreover, the proposed all-dielectric metasurface based color filters can be integrated with the well matured fabrication technology of electronic devices.
\end{abstract}


With tremendous changes in nanotechnology over past few decades, it becomes possible to fabricate devices which promise to revolutionize many areas. Examples include ultra-thin planar lens~\cite{kildishev2013planar,yu2014flat}, optical sensing~\cite{adato2009ultra,liu2010infrared}, photo-voltaic devices~\cite{atwater2010plasmonics,chen2012broadband}, non-fading colors~\cite{james2016plasmonic}, and various holography based devices~\cite{almeida2015nonlinear,ni2013metasurface}. In particular, color pixels using nanoparticles have gained significant attention in recent years because of several advantages over pigment based color printing techniques
like high resolution~\cite{tan2014plasmonic}, high contrast, everlasting colors, significant low power consumption, and recyclability of  product~\cite{clausen2014plasmonic}. The concept of structural color printing is inspired by observations in nature, such as morpho butterflies, beetles, and the feathers of peacocks~\cite{srinivasarao1999nano,vukusic2001structural,kinoshita2008physics,gralak2001morpho}. However, these colors are highly sensitive to the variations in the angle of incidence, shape, and size of the nanostructure. 
To make this plasmonics based structural technology 
more mature, its angle dependency~\cite{wu2013angle,hojlund2014angle}, sensitivity to polarization, and ease of fabrication must be taken into account. In recent years, many efforts have been done to study the aforementioned issue in plasmonic color printing~\cite{kumar2012printing,roberts2014subwavelength,ellenbogen2012chromatic,diest2013aluminum,li2016dual,duempelmann2015color,wang2016large,richner2016full,fan2017three}. Earlier, the most commonly used materials for plasmonic nanostructure based pixels have been gold and silver~\cite{ehrenreich1962optical,west2010searching}. Gold has interband transition in the lower visible regime~\cite{ehrenreich1962optical}, while silver is suitable for the entire visible range but is susceptible with the native oxide that spoils the stability of colors. Moreover, gold and silver are not economical for large scale integration. Aluminum is probably the most prominent candidate~\cite{gerard2014aluminium}. It is more robust and economical for 
large-scale fabrication~\cite{james2016plasmonic}. However, it shows lower quality ($i.e.$, broader) resonance in the visible spectrum than gold or silver, especially at 800{\nano\meter} wavelength, where interband transition takes place. Ultimately,  all these metal based plasmonic devices show significant losses within the visible spectrum.\\ 

On the other hand, all-dielectric metasurfaces can be a promising solution with significant advantages over metallic nanostructures such as high quality resonances and low intrinsic ohmic losses~\cite{paniagua2016generalized,bonod2015silicon,jahani2016all,decker2015high,sautter2015active,li2016all,moitra2015large,liu2014optical,hegde2016design}. Silicon based all-dielectric devices have been reported for local manipulation by wavefronts, such as beam diversion, vortex plates and light focusing using meta-lenses~\cite{shalaev2015high,decker2015high,doi:10.1021/nn402736f,lin2014dielectric,arbabi2015dielectric}. The advantages of $Si$ nanodisks are high refractive index and ease of fabrication with well established CMOS technology. Interestingly, the high refractive index allows to manipulate by magnetic and electric components of light simultaneously. In the case of metal based nanoantennas, absorption losses can be significant at visible spectrum, while interaction with magnetic component of the incident beam requires more complex shapes. Recently, an investigation has been conducted to demonstrate the possibility of using silicon-aluminum hybrid nanodisks~\cite{yue2016subtractive,shrestha2015polarization} to create colors of high quality. Silicon nanoparticles were proposed as a valuable alternative to plasmonic nanoantennae for the design of color pixels~\cite{cao2010tuning, proust2016all,zhao2016full,hegde2016design}. However, the potential 
of all-dielectric resonance structures is presently very far from being fully estimated and exploited. \\

In this work, we propose a systematic approach to build color filters by using advantages of cross-shaped $Si$ nanoresonators, which are closely spaced to each other to create a metasurface. Recently reported numerical studies of the nanocross geometry~\cite{hegde2016design} have indicated that a broader gamut of colors is possible in comparison to simpler shapes like the cylinder (disk). The main goal is to obtain a high quality (narrow) resonance throughout the visible spectrum that enables an extended gamut with colors of high purity. It is known that $Si$ nanostructures of different shapes typically offer an opportunity to excite individual electric type and magnetic type Mie resonances, or both resonances simultaneously~\cite{kuznetsov2012magnetic}. In fact, it has been demonstrated that by tuning the aspect ratio carefully, one can overlap both resonances to achieve near unity transmission~\cite{doi:10.1021/nn402736f}. In this paper, the all-dielectric metasurfaces are used in reflection mode. A very confined energy is concentrated within the structure due to the high quality of the used Mie resonances.\\

The main hypothesis that we follow here is based on the expectation that a proper manipulation by the selected Mie resonances may enable desired improvements of the resulting resonance quality owing to better confinement of resonance fields and, simultaneously, removal of secondary (unwanted) spectral
features, so that enrichment of colors can be achieved. We decided in favor of cross-shaped $Si$ nanoresonators as building elements, which are expected to be suitable~\cite{hegde2016design} for achievement of the goals of this study. Each of them is made of two identical orthogonal rectangle-shaped $Si$ nanoantennas. In this case, resonances are governed by cross-shaped nanoantennas and thus, colors can be controlled $via$ all three geometrical parameters of individual nanoantennas. This gives a new degree of freedom as compared to the nanodisks, that is highly demanded for efficient optimization. Using the suggested approach, we predict by simulations and confirm experimentally that one can easily achieve a high quality resonance for the entire visible spectrum by carefully choosing the length and width of the cross-shaped nanoresonators.

\section{Results}
\begin{figure}[!htb]
\centering
\includegraphics[scale=0.85]{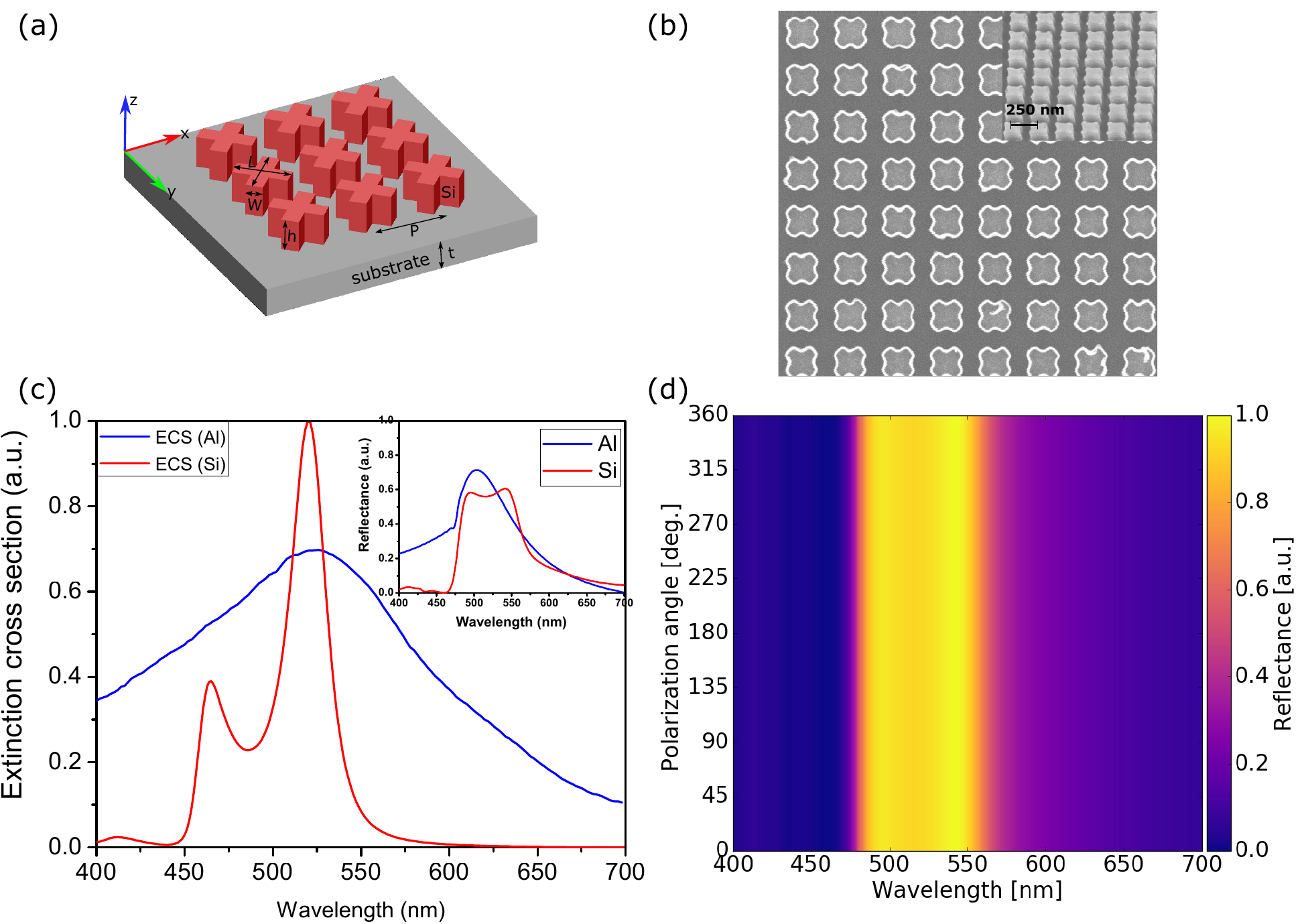}
\caption{\textbf{Perspective view and SEM images of the all-dielectric metasurface, extinction cross section (ECS) spectra and reflectance spectra, and reflectance \textit{vs} polarization angle.} (a) Schematic representation of the array of cross-shaped $Si$ nanoresonators on top of the quartz substrate. The thickness of the substrate $t$ = 275{\micro\meter}. For each nanoantenna, height $h$ = 140{\nano\meter}, length $L$ and width $W$ are scaled to achieve different colors. The center-to-center distance between the two nanoresonators (lattice constant) is $P = 250{\nano\meter}$. (b) Top view of SEM images of the fabricated structure with $P$ = 250{\nano\meter}. A 45$^{\circ}$ cross section view is added in the inset. (c) ECS spectra in case of $Si$ and $Al$ nanoantennae. Two peaks arising in the former case are due to electric type and magnetic type resonance (see supplementary information for the field patterns), while 
there is only single broad resonance in the latter case. The inset shows the reflectance spectra for the same two structures.
The length, width and height are 100{\nano\meter}, 50{\nano\meter} and 140{\nano\meter}, respectively. 
(d) A colormap of simulated reflectance spectra of the $Si$ based metasurface at polarization angle varied from 0$^{\circ}$ to 360$^{\circ}$.}
\label{main_figure}
\end{figure}
Let us start from the general geometry and basic operation principles of the proposed devices. Figure~\ref{main_figure}(a) presents the perspective view of the proposed all-dielectric metasurface together with some details of geometry. The cross-shaped $Si$ nanoresonators are deposited on top of the quartz substrate (see Methods of fabrication). The height of nanoantennae is selected as 140{\nano\meter} (in subwavelength range). Figure~\ref{main_figure}(b) represents the top view of SEM image of the device. A 45$^{\circ}$ cross section view is also added in the inset for the same fabricated device. For the studied $Si$ structure, extinction cross section spectrum is presented in Fig.~\ref{main_figure}(c). 
Two resonance peaks are observed 
at 465{\nano\meter} and 520{\nano\meter}. 
They can be tuned throughout the visible range by changing the length-to-width aspect ratio of individual rectangle-shaped nanoantennas. 
The $Si$ nanoresonator dimensions have been optimized to excite these two resonances as close as possible but without a full overlapping. In addition, the criterium of minimizing unwanted spectral features has been applied in order to obtain more gradual behavior in the working spectral range. 
As follows from the obtained simulation results, optimization yields a resonance range that is narrower and, thus, corresponds to a resonance of higher quality, as compared to the case of $Al$ cross-shaped nanoantennae, see Fig.~\ref{main_figure}(c). 
We have also compared the simulated reflectance spectra for the metal and $Si$ based structures at the same dimensions [see Fig.~\ref{main_figure}(c), inset]. These results confirm that the metal nanostructure features broader resonances than the engineered $Si$ one. An important advantage of cross-shaped nanoantennae is that they preserve the polarization independence. As an example, Fig.~\ref{main_figure}(d) presents the simulated reflectance spectrum for the entire range of polarization angle variation and entire wavelength range considered. The obtained results confirm that there is no change in the reflectance spectrum when the polarization angle is varied. 

\begin{figure}[!htb]
\centering
\includegraphics[width=\linewidth]{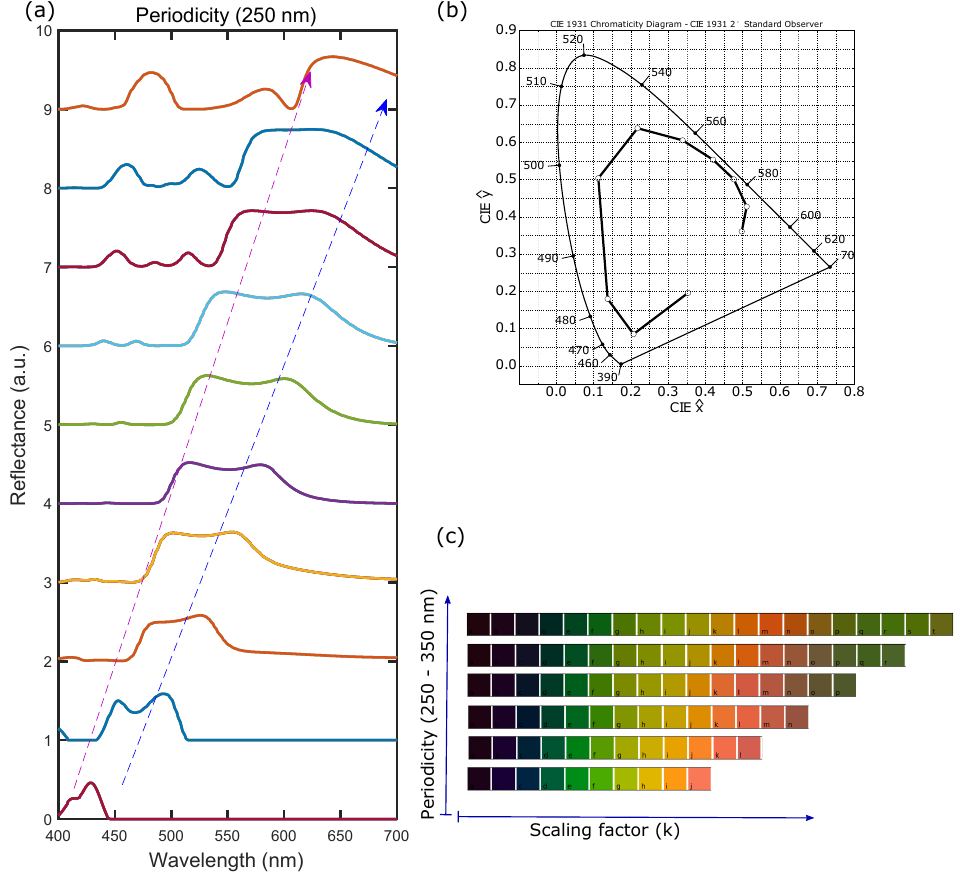}
\caption{\textbf{Reflectance spectra (simulation), corresponding chromaticity diagram, and photograph of experimental images of the array visible under optical microscope.} (a) Unit-cell simulation results for cross shaped $Si$ nanoresonators on quartz substrate; the initial values of geometrical parameters are $L$ = 65{\nano\meter}, $W$ = 35{\nano\meter}, and $P$ = 250{\nano\meter}. $L$ and $W$ are linearly scaled from 65{\nano\meter} to 195{\nano\meter} and 35{\nano\meter} to 105{\nano\meter}, respectively, from bottom to top.
The resonances are redshifted in the visible regime, as schematically shown by arrows. (b) Representation of reflectance spectra on standard CIE 1931 chromaticity diagram for $P$ = 250{\nano\meter}. (c) Experimental colors visible under optical microscope for different values of $P$ which are varied from 250{\nano\meter} (the lowest series) to 350{\nano\meter} (the most upper series) with step of 20{\nano\meter} (from the lowest series to the most upper one); $L$ and $W$ are linearly scaled from 65{\nano\meter} to 260{\nano\meter} and 35{\nano\meter} to 140{\nano\meter}, respectively; with $k=L/\mbox{$\min$}(L)=W/\mbox{$\min$}(W).$}
\label{all_three}
\end{figure}

Since a specific color results from resonant interaction of light with nanoresonators, it can be obtained from adjustment of geometrical parameters that properly affect spectral locations and properties of Mie resonances. The possibility of obtaining multiple colors with the aid of metasurfaces like that in Fig.~\ref{main_figure}(a) and (b) is demonstrated in Fig.~\ref{all_three}.
The length and width of rectangle-shaped $Si$ nanoantennae are simultaneously linearly scaled in order to tune the electric and magnetic type resonances in the entire visible spectrum from 400{\nano\meter} to 700{\nano\meter}, as shown in Fig.~\ref{all_three}(a) for $P=250${\nano\meter}. 
A commercial-grade simulator based on the finite-difference time-domain method~\cite{Lumerical} is used to perform the calculations.
They are conducted for a unit cell with periodic boundary conditions, and varied lattice constant from 250{\nano\meter} to 350{\nano\meter}, by keeping the periodicity in the subwavelength range (see Methods, Simulation).
Each spectral zone in Fig.~\ref{all_three}(a) corresponds to a specific color. It is clearly seen that the electric and magnetic type resonances can be tuned through the entire visible wavelength spectrum, as desired. Conversion of reflectance spectra into colors on CIE1931 chromaticity diagram can be performed, in the general case, by using an open source Python program~\cite{colour_science}. The results of conversion of the spectra shown in Fig.~\ref{all_three}(a) are presented in Fig.~\ref{all_three}(b). Generally, a higher quality of resonances corresponds to a better approaching to the boundaries of the chromaticity diagram and, hence, enable higher quality and wider gamut of colors. Complete details about color visualization using reflectance spectra are given in supplementary information under section color representation from reflectance spectra. 

By operating the metasurface in reflection mode, a broad spectrum of colors for highly selective wavelengths ($i.e.$, high quality colors) can be obtained. In principle, colors can be generated by using 
either additive or subtractive approach~\cite{colour_approach}. Here, we have used the additive approach. Ideally, the reflection spectrum must be as narrow as possible in order to generate a very specific color. A narrower resonance represents a more specific wavelength color, whereas the amplitude of the peak decides the saturation level of the color. With the aid of high quality narrow resonances, we improve the approaching to the boundaries of CIE-1931 chromaticity diagram, so a color of higher quality and a wider gamut of colors
can be obtained, as desired. We experimentally found that different colors can be obtained at different values of period ($P$, lattice constant), which  correspond to the scaled length $(L)$ and width $(W)$ of the nanoantenna, see Fig.~\ref{all_three}(c).
Each square in Fig.~\ref{all_three}(c) corresponds to a unique set of geometrical parameters. The lowest series of the squares shown here corresponds to the structures, for which reflectance spectra are presented in Fig.~\ref{all_three}(a).
Thus, the resonance region corresponds to different colors at different values of $P$, see Fig.~S4 in supplementary information.
This dependence occurs owing to the coupling of resonance fields of nanoresonators. 
The use of larger values of $P$ allows us to create a richer variety of colors, as we have more choices to increase the length and width. We have observed different colors under optical microscope due to variations in lattice constant $(P)$ from 250{\nano\meter} to 350{\nano\meter}, see Fig.~\ref{all_three}(c). 
The lattice constant was increased here by a reasonable increment of 20{\nano\meter} to make it feasible for fabrication process. Although it might be hard to distinguish between the highly saturated colors in Fig.~\ref{all_three}(c), the reflectance spectra in Fig.~\ref{all_three}(a) 
and the corresponding CIE-1931 chromaticity diagram in Fig.~\ref{all_three}(b) give us a clear picture about it. In fact, a color gamut can be possible by making a matrix between the scaled lengths and widths.

The fact that two resonances, which are observed in Fig.~\ref{all_three}(a) at different values of $L$ and $W$, are closely spaced makes fabrication of a particular color possible, that is unlikely in case of metal based plasmonic structures, because they show a broad resonance. Moreover, it is possible to create a selective wavelength color due to sharp resonances, particularly in the lower part of the visible spectrum. 
It is observed in Fig.~\ref{all_three}(a) that as we increase the size of cross-shaped resonators some additional Mie resonances are also excited, in coincidence with the predictions based on the simulation results. These resonances reduce the hue and saturation of red color, because of mixing contribution of different frequencies. So the red color seems to be the most difficult one to fabricate. Below, we will show that in spite of the above-mentioned difficulties the suggested structure allows us creating fairly red colors by carefully adjusting the values of $P$, $L$, and $W$. Thanks to this adjustment, the unwanted effect of higher-order resonances can be minimized. \\

\begin{figure}[!htb]
\centering
\includegraphics[width=\linewidth]{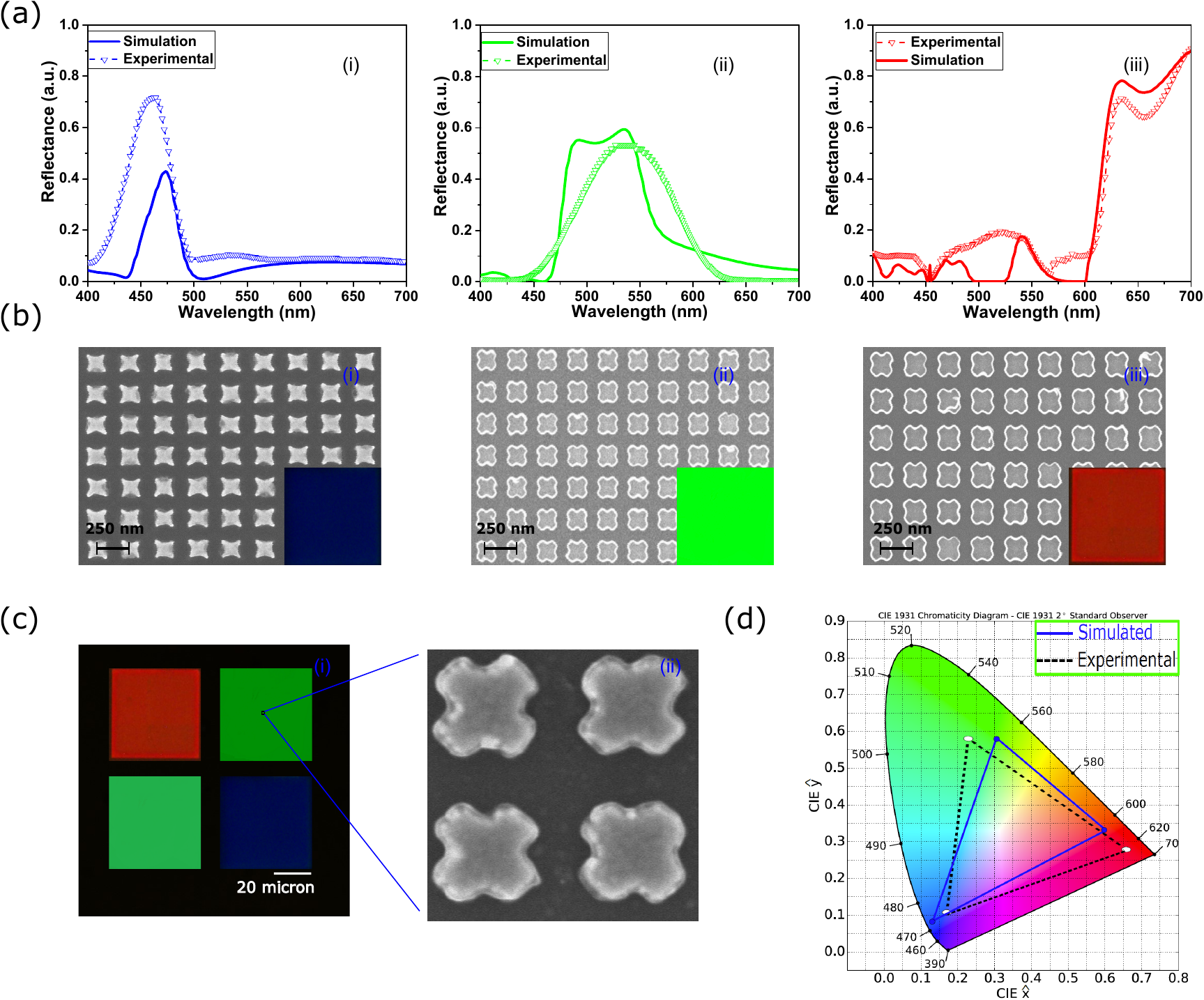}
\caption{\textbf{Simulation and experimental results for primary colors with SEM images.} (a) Simulated and experimental reflectance spectra. The dimensions of cross-shaped $Si$ nanoresonators (i) $L$ = 85{\nano\meter} and $ W$ = 46{\nano\meter} for blue color. (ii) $L$ = 114{\nano\meter} and $W$ = 62{\nano\meter} for green color and (iii) $L$ = 215{\nano\meter} and $ W$ = 116{\nano\meter} for red color. The lattice constant is 250{\nano\meter} for all three cases. (b) SEM images of the fabricated structures with the inset view of associated colors. (c) A photograph taken from Nikon camera attached with 50$\times$ lens with $NA = 0.65$ and the enlarged SEM image of the nanostructure. (d) A CIE 1931 chromaticity diagram that is used to visualize the simulated and measured colors.}
\label{rgb_spectra}
\end{figure}

The three primary colors (RGB) represent the fundamental unit for color printing technology. All the other colors in the RGB gamut can be derived by mixing the primary colors appropriately. Figure \ref{rgb_spectra} presents the results of a detailed experimental demonstration of the suggested devices in the form of pixels. 
A dual characterization is done to ensure the results by measuring the reflectance spectra of the samples with the aid of a home-made customized setup and observing the colors directly under optical microscope (see Methods, Optical characterization). Figure~\ref{rgb_spectra}(a) shows the experimental and simulated reflectance spectra for highly saturated primary colors. These results show good agreement with each other. Figure~\ref{rgb_spectra}(b) shows the SEM images obtained at different sizes of nanoantennas. Insets are added to the SEM images to show the corresponding colors visible under optical microscope, which are associated with the different sizes of the cross-shaped nanoresonators. Finally, these three primary colors are fabricated in a form of pixel, being the main component of any display device. The size of each square block is 50{\micro\meter}. Details of the used fabrication method are given at the end of the paper. The optical microscope images shown in Fig.~\ref{rgb_spectra}(c) confirm the quality of highly saturated primary colors, which is an important advantage of the suggested all-dielectric metasurface based pixels over the existing plasmonics devices. A CIE 1931 chart is used to represent the simulated and experimental spectra of the primary colors, see Fig. 3(d). One can see a  very small shift in color spectrum, which might come from fabrication imperfections. 
It is noticeable that there is good coincidence between two sets of experimental results.

\section{Conclusion}
Polarization insensitive all-dielectric metasurfaces based on 2D arrays of cross-shaped $Si$ nanoresonators have been proposed to realize color filters with extended gamut for the entire visible spectrum. A numerical investigation has been carried out that demonstrates the principal possibility of obtaining high-purity colors by means of optimization of resonance properties, which can be realized by a relatively simple adjustment of the structural parameters. The role of existence and properties of the dual resonance, which is achieved at a partial overlapping of electric type and magnetic type resonances, and that of suppression of unwanted spectral features in the obtaining of these advancements have been clarified. The utilized resonances can be tuned by changing the length-to-width aspect ratio of individual rectangle-shaped nanoantennas. This concept has been used to design and fabricate the color filters. Our simulation results reasonably agree with the experimental ones. Some differences should be noticed that may be connected with fabrication complexity of the structure. The experimentally demonstrated possibility of obtaining high quality (narrow) resonances, which enable high quality colors, is the most important result of this work. We have demonstrated the wide variety of colors for different periodicity and size of 
cross-shaped nanoresonators. Additionally, we have demonstrated the primary colors painting in the form of pixels.
These colors show high saturation and hue value. In fact, our device is capable to produce a large panel color in the visible regime with strong spectral selectivity, provided that the nanoantenna aspect ratio is properly chosen. By carefully controlling the balance between desired and unwanted Mie resonances, one can further optimize the color filter, especially in dark zone of red color. Since the $a-Si$ (amorphous-$Si$) is the most suitable material for large-scale fabrication with the existing technology, it can potentially be used for making low-cost, eco-friendly, high quality, long lasting painting possible for mass production in the future.
\section{Methods}
\textbf{Simulations.} We have used Lumerical FDTD solver~\cite{Lumerical} to study metasurfaces comprising the cross-shaped nanoresonators on a dielectric substrate. The materials used for substrate and cross-shaped nanoresonators are $SiO_2$ and $Si$, respectively. The material parameters are taken from default the material library of the used software. A plane wave ranging from 400{\nano\meter} to 700{\nano\meter} is illuminated from the top of the structure. Periodic boundary conditions are used in the unit cell along $x$ and $y$ directions. Perfect matching layer (PML) boundary conditions were used in the $z$ directions to avoid any reflection. The reflectance spectra are simulated by considering a unit cell (one cross-shaped nanoresonator on substrate) with periodic boundary conditions in $x$ and $y$ directions.\\
\textbf{Device fabrication.} A piranha cleaned quartz sample (275{\micro\meter} thick) is used to fabricate the device. We have deposited a thin layer of 140{\nano\meter} amorphous $Si$ using ICPCVD tool at 300\degree Celsius with 150W added microwave power. A single-layer PMMA photoresist is used for patterning cross-shaped nanoresonators by using Raith 150-Two EBL tool. An electronic mask is designed using an open source Python program. The exposed sample is developed using MIBK-IPA (1:3) and an IPA solution for 45{\second} and 15{\second}, respectively. A thin layer of metal (5{\nano\meter} \textit{Cr} as adhesion layer and 40{\nano\meter} \textit{Au}) is deposited to transfer the pattern on metal layer for lift-off process using four target evaporators. After lift-off, the sample is etched using plasma asher to get the final pattern. A process flow chart with step by step details is available in supplementary information. \\
\textbf{Optical characterization.} A dual optical characterization is done to ensure the results. The sample is placed under Olympus optical microscope and illuminated with white light without filter. The colors can be directly seen under optical microscope. The reflectance spectra are measured using a home-made customized setup. A HL 2000 halogen lamp source is coupled with an optical fiber to illuminate the sample in the visible range, $i.e.$, from 400{\nano\meter} to 700{\nano\meter}. A 50$\times$ objective lens with $NA = 0.65$ is used to get tight focusing of light on the sample. The reflectance spectra are measured using the same objective lens. All the collected data are normalized with respect to the bare quartz sample. A Nikon camera attached with assembly is used to take the photograph of the illuminated area.

\begin{acknowledgement}
RSH acknowledges support from Department of Science and Technology, India under the Extramural Research Grant no. SB/S3/EECE/0200/2015. RSH, VV and GV acknowledge support from Indian Nanoelectronics Users Program under grant nos. P643987963 and P875860276. The work was partially supported by the National Science Centre Poland for OPUS grant No. 2015/17/B/ST3/00118(Metasel) and by the European Union Horizon2020 research and innovation program under the Marie Sklodowska-Curie grant agreement No. 644348 (MagIC). Authors thank all the members of CEN laboratory, IIT Bombay who helped us directly or indirectly while doing nanofabrication work. Special thanks to Dr.\ K Nageshwari and Dr.\ Ritu Rashmi for providing necessary facilities and regular advice.
\end{acknowledgement}

\begin{suppinfo}

\end{suppinfo}

\bibliography{achemso-demo}



\end{document}